\documentclass[twocolumn]{aastex62}
\graphicspath{{./}{figures/}}
\usepackage{verbatim}
\usepackage{hyperref}
\usepackage{amsmath}
\usepackage{graphicx}
\usepackage[normalem]{ulem}
\shorttitle{Pulsar Planet Constraints}
\shortauthors{Behrens et al.}
\begin{document}

\title{The NANOGrav 11-year Data Set: Constraints on Planetary Masses Around 45 Millisecond Pulsars}

\author[0000-0002-2333-5474]{E. A. Behrens}
\affil{Department of Astronomy, The Ohio State University, 4055 McPherson Laboratory, 140 West 18th Ave, Columbus, OH 43210, USA}

\author[0000-0001-5799-9714]{S. M. Ransom}
\affil{National Radio Astronomy Observatory, 520 Edgemont Rd., Charlottesville, VA 22903, USA}

\author[0000-0003-2285-0404]{D. R. Madison}
\altaffiliation{NANOGrav Physics Frontiers Center Postdoctoral Fellow}
\affil{Department of Physics and Astronomy, West Virginia University, P.O. Box 6315, Morgantown, WV 26506, USA}
\affiliation{Center for Gravitational Waves and Cosmology, West Virginia University, Chestnut Ridge Research Bldg, Morgantown, WV 26505, USA}

\author{Z. Arzoumanian}
\affiliation{X-Ray Astrophysics Laboratory, NASA Goddard Space Flight Center, Code 662, Greenbelt, MD 20771, USA}









\author{K. Crowter}
\affiliation{Department of Physics and Astronomy, University of British Columbia, 6224 Agricultural Road, Vancouver, BC V6T 1Z1, Canada}

\author[0000-0002-2185-1790]{M. E. DeCesar}
\altaffiliation{NANOGrav Physics Frontiers Center Postdoctoral Fellow}
\affiliation{Department of Physics, Lafayette College, Easton, PA 18042, USA}

\author[0000-0002-6664-965X]{P. B. Demorest}
\affiliation{National Radio Astronomy Observatory, 1003 Lopezville Road, Socorro, NM 87801, USA}

\author[0000-0001-8885-6388]{T. Dolch}
\affiliation{Department of Physics, Hillsdale College, 33 E. College Street, Hillsdale, Michigan 49242, USA}

\author{J. A. Ellis}
\affiliation{Infinia ML, 202 Rigsbee Avenue, Durham NC, 27701}

\author[0000-0002-2223-1235]{R. D. Ferdman}
\affiliation{School of Chemistry, University of East Anglia, Norwich, NR4 7TJ, United Kingdom}

\author{E. C. Ferrara}
\affiliation{NASA Goddard Space Flight Center, Greenbelt, MD 20771, USA}

\author[0000-0001-8384-5049]{E. Fonseca}
\affiliation{Department of Physics, McGill University, 3600  University Street, Montreal, QC H3A 2T8, Canada}


\author[0000-0001-8158-638X]{P. A. Gentile}
\affiliation{Department of Physics and Astronomy, West Virginia University, P.O. Box 6315, Morgantown, WV 26506, USA}
\affiliation{Center for Gravitational Waves and Cosmology, West Virginia University, Chestnut Ridge Research Bldg, Morgantown, WV 26505, USA}





\author{G. Jones}
\altaffiliation{NANOGrav Physics Frontiers Center Postdoctoral Fellow}
\affiliation{Department of Physics, Columbia University, New York, NY 10027, USA}

\author[0000-0001-6607-3710]{M. L. Jones}
\affiliation{Center for Gravitation, Cosmology and Astrophysics, Department of Physics, University of Wisconsin-Milwaukee,\\ P.O. Box 413, Milwaukee, WI 53201, USA}


\author[0000-0003-0721-651X]{M. T. Lam}
\affil{School of Physics and Astronomy, Rochester Institute of Technology, Rochester, NY 14623, USA}
\affil{Laboratory for Multiwavelength Astronomy, Rochester Institute of Technology, Rochester, NY 14623, USA}


\author[0000-0002-2034-2986]{L. Levin}
\affiliation{Department of Physics and Astronomy, West Virginia University, P.O. Box 6315, Morgantown, WV 26506, USA}
\affiliation{Center for Gravitational Waves and Cosmology, West Virginia University, Chestnut Ridge Research Bldg, Morgantown, WV 26505, USA}


\author[0000-0003-1301-966X]{D. R. Lorimer}
\affiliation{Department of Physics and Astronomy, West Virginia University, P.O. Box 6315, Morgantown, WV 26506, USA}
\affiliation{Center for Gravitational Waves and Cosmology, West Virginia University, Chestnut Ridge Research Bldg, Morgantown, WV 26505, USA}


\author[0000-0001-5229-7430]{R. S. Lynch}
\affiliation{Green Bank Observatory, P.O. Box 2, Green Bank, WV 24944, USA}


\author[0000-0001-7697-7422]{M. A. McLaughlin}
\affiliation{Department of Physics and Astronomy, West Virginia University, P.O. Box 6315, Morgantown, WV 26506, USA}
\affiliation{Center for Gravitational Waves and Cosmology, West Virginia University, Chestnut Ridge Research Bldg, Morgantown, WV 26505, USA}



\author[0000-0002-3616-5160]{C. Ng}
\affiliation{Dunlap Institute for Astronomy and Astrophysics, University of Toronto, 50 St. George Street, Toronto, ON M5S 3H4, Canada}

\author[0000-0002-6709-2566]{D. J. Nice}
\affiliation{Department of Physics, Lafayette College, Easton, PA 18042, USA}

\author[0000-0001-5465-2889]{T. T. Pennucci}
\altaffiliation{NANOGrav Physics Frontiers Center Postdoctoral Fellow}
\affil{National Radio Astronomy Observatory, 520 Edgemont Rd., Charlottesville, VA 22903, USA}
\affiliation{Institute of Physics, E\"{o}tv\"{o}s Lor\'{a}nd University, P\'{a}zm\'{a}ny P. s. 1/A, 1117 Budapest, Hungary}

\author[0000-0002-8509-5947]{B.~B.~P.~Perera}
\affiliation{Arecibo Observatory, HC3 Box 53995, Arecibo, PR 00612, USA}

\author[0000-0002-5297-5278]{P. S. Ray}
\affiliation{Space Science Division, Naval Research Laboratory, Washington, DC 20375-5352, USA}


\author[0000-0002-6730-3298]{R. Spiewak}
\affiliation{Centre for Astrophysics and Supercomputing, Swinburne University of Technology, P.O. Box 218, Hawthorn, Victoria 3122, Australia}

\author[0000-0001-9784-8670]{I. H. Stairs}
\affiliation{Department of Physics and Astronomy, University of British Columbia, 6224 Agricultural Road, Vancouver, BC V6T 1Z1, Canada}


\author[0000-0002-7261-594X]{K. Stovall}
\altaffiliation{NANOGrav Physics Frontiers Center Postdoctoral Fellow}
\affiliation{National Radio Astronomy Observatory, 1003 Lopezville Road, Socorro, NM 87801, USA}

\author[0000-0002-1075-3837]{J. K. Swiggum}
\altaffiliation{NANOGrav Physics Frontiers Center Postdoctoral Fellow}
\affiliation{Department of Physics, Lafayette College, Easton, PA 18042, USA}





\author[0000-0001-5105-4058]{W. W. Zhu}
\affiliation{National Astronomical Observatories, Chinese Academy of Science, 20A Datun Road, Chaoyang District, Beijing 100012, China}
 
\noaffiliation

\begin{abstract}
We search for extrasolar planets around millisecond pulsars using pulsar timing data and seek to determine the minimum detectable planetary masses as a function of orbital period. Using the 11-year data set from the North American Nanohertz Observatory for Gravitational Waves (NANOGrav), we look for variations from our models of pulse arrival times due to the presence of exoplanets. No planets are detected around the millisecond pulsars in the NANOGrav 11-year data set, but taking into consideration the noise levels of each pulsar and the sampling rate of our observations, we develop limits that show we are sensitive to planetary masses as low as that of the moon. We analyzed potential planet periods, $P$, in the range $7~{\rm days}<P<2000~{\rm days}$, with somewhat smaller ranges for some binary pulsars. The planetary mass limit for our median-sensitivity pulsar within this period range is $1~M_{\rm moon}(P/100~{\rm days})^{-2/3}$. 
\end{abstract}

\section{Introduction} \label{sec:intro}

In 1992, the first two confirmed exoplanets were found around the isolated pulsar B1257+12 using pulsar timing techniques \citep{wf92}. This 6.2-ms pulsar was initially found to host two planets, each being a few earth masses with periods of 2-3 months. Two years after this initial discovery, another periodic signal was discovered and later confirmed to be a third exoplanet around the same pulsar, and to this day, this planet is still the least massive confirmed exoplanet \citep{w94}, with a mass of only about two moon masses, or about 10$^{23}$ kg. In 2000, a circumbinary planet was discovered around PSR B1620$-$26 and its white dwarf companion, and this planet was also announced to be the oldest planet discovered with an age of about 12.6 Gyr \citep{fjr+00}. 

Methods of pulsar planet detection are described in \cite{k18}, and early work in the field of pulsar planets was reviewed in \cite{pt94}. There have been few recent systematic searches for pulsar planets, particularly around millisecond pulsars. \cite{k+15} search for planets around a large number of pulsars, but they were non-millisecond pulsars. \cite{c+18} used millisecond pulsar data to search for unknown planets in our own Solar System, but their method, cross-correlating data across pulsars, precluded detection of planets around individual pulsars.

In total, six planetary-mass bodies have been confirmed to be orbiting pulsars \citep{bbb+11,sbb+18}, but the mechanism of forming pulsar-planet systems is unclear. If these planets formed in the protoplanetary disc phase, we would not expect them to survive the subsequent supernova explosion, making the discovery of pulsar planets a somewhat unexpected result. One theory proposes that when the supernovae that produce these pulsars explode, some of the matter that is ejected is then gravitationally recaptured by the pulsar and forms a rotating disk, functioning similarly to the protoplanetary disks with which we are familiar \citep{lwb91}.

Another theory that has been proposed for pulsars in binary orbits involves the planet matter actually coming from the pulsar's companion. Many millisecond pulsars (MSPs) that are thought to have been spun up to their current rotational period by accreting matter from a low-mass donor star and receiving an influx of angular momentum eventually vaporize their companions, earning the nickname ``black widow" pulsars \citep{r11,bv91,acw99}. It has been suggested that the high-energy relativistic wind from these pulsars could ablate their low-mass companions, but the matter being chipped away is then recaptured by the pulsar and forms a rotating disk \citep{p10}. Like the previous model, planetesimals would then form from this disk similarly to how they form in protoplanetary disks around newly-formed stars. It is even possible that mechanisms in the theories described above could lead to the formation of asteroid belts around pulsars \citep{scm+13}. 

In this paper, we discuss and analyze our search for planets around MSPs in the 11 year data set \citep{abb+18} from the North American Nanohertz Observatory for Gravitational Waves (NANOGrav). NANOGrav is a collaboration of scientists from the United States and Canada whose goal is to use pulsars as a galactic-scale detector for long-period gravitational waves \citep{bcc+19}. This data set contains approximately 11 years of time of arrival (TOA) data for 45 MSPs and their timing ephemerides. In the pursuit of gravitational wave detection, NANOGrav has compiled a data set of some of the most highly stable MSPs in the sky with timing precision of under 1-2 $\mu$s, which makes this data set an excellent resource for not only gravitational wave studies, but for other astrophysics as well. 

This paper describes our use of the NANOGrav 11-year data set to search for exoplanets around pulsars as well as to set lower limits on the masses of exoplanets that could be detected using this data set. In Section \ref{sec:data}, we expand upon the qualities of the NANOGrav data set. In Section \ref{sec:sig} we discuss how we expect to see planetary signals in pulsar timing residuals (the difference between observed and predicted TOAs) and how we optimize the data set for exoplanet searches. In Section \ref{sec:techniques} we introduce our detection scheme as well as our methods for determining mass constraints. In Section \ref{sec:results} we present the results of our search and our efforts to define detection limits, and finally in Section \ref{sec:conc} we summarize our work and propose some future work in this field.

\section{Data} \label{sec:data}
The NANOGrav 11-year data set contains high-precision TOA measurements as well as timing models for 45 MSPs taken over the span of roughly 11 years, though observations of some pulsars span longer or shorter periods of time. Observations of these pulsars were taken with a roughly monthly, though sometimes weekly, cadence on the 305-m William E. Gordon Telescope at the Arecibo Observatory as well as on the 100-m Robert C. Byrd Green Bank Telescope at the Green Bank Observatory. Of these 45 pulsars, 31 are in binary systems, the orbits of which we have modeled with both Keplerian and post-Keplerian parameters to account for apparent orbital deviations from Keplarian motion \citep{d&t92}.

In this paper, we primarily consider the pulsar timing residuals of the NANOGrav data to analyze the variation between the data and our models. These models were created using a weighted, linear least-squares fit generated using {\tt TEMPO} \citep{ascl15}, a software package commonly used in pulsar timing. For most of the pulsars included in this data set, the root-mean-square of the timing residuals is below 2 $\mu$s; the only exceptions include those pulsars for which the collected data exhibit strong red noise, or low-frequency noise \citep{abb+18}.

\section{Planetary Signals in Pulsar Timing Data} \label{sec:sig}
To determine if exoplanets are orbiting the MSPs in this data set, we analyze the timing residuals to search for a characteristic variation we would expect to see from our models as a result of a planetary presence. Due to the reflex motion of a pulsar in response to a planetary orbit, we expect to see fluctuations in pulsar TOAs as the pulsar orbits the center of mass (COM) of the pulsar-planet system. The pattern of these fluctuations would repeat periodically and, assuming a circular orbit, would therefore be seen as a sinusoidal structure in the pulsar's residuals. By extracting the amplitude and frequency of these sinusoids, we can gather information about the pulsar's distance from the COM and the frequency of its orbits, which is also equal to the planet's orbital frequency. We can input these parameters into the equations for Kepler's third law 
\begin{equation}
\label{eq:Kep3}
P^2 = \frac{4\pi^2\left(R + r\right)^3}
{G\left(M + m\right)}
\end{equation}
and the definition of the COM 
\begin{equation}
\label{eq:COM}
M  R = m r,
\end{equation}
where $P$ is the planet's orbital period; $R$ and $r$ are the distances to the system's COM from the pulsar (and its companion, if present) and the planet, respectively; $M$ is the central mass of the pulsar system, either the pulsar's mass $M_1$ by itself in the case of an isolated pulsar or the combined mass of the pulsar and the mass of its companion $M_1 + M_2$; and $m$ is the mass of the planet. We use these equations to ultimately calculate the mass of the planet associated with the aforementioned sinusoidal signal and find the relationship $m \propto P^{-2/3}$.

To determine some characteristics of the planetary signals we might see in our data, we carefully consider the orbital frequencies that would be detectable and would make the most physical sense given the parameters of our systems. As a result of the, at most, weekly sampling of the NANOGrav data, we determine that we would be unlikely to detect any planets with orbital periods of less than one week, so we define 7 days as our generic minimum orbital period to be probed. For the pulsars in binary orbits, additional analysis is done to account for the various possible planetary configurations given the characteristics of our binaries as described below.

We consider two possible planetary configurations for the binary systems in the NANOGrav 11-year release, both of which are described in detail in \citet{cqd+02}. The first configuration we consider is a planet in a satellite S-type orbit, in which the planet orbits only the pulsar, and the companion orbits both the planet and the pulsar around the outskirts of the system. \citet{hw99} define a critical semimajor axis $a_{c}$ for planets in this configuration in which planets with semimajor axis $ a \leq a_{c}$ will be stable against perturbations caused by the binary companion. For S-type orbits, the semimajor axes follow the relationship
\begin{align}
\label{eq:stype}
a \leq [&(0.464\pm0.006)+
(-0.380\pm0.010)\mu+ \nonumber\\
&\left(-0.631\pm0.034\right)e+
\left(0.586\pm0.061\right)\mu e + \nonumber\\
&\left(0.150\pm0.041\right)e^2+
\left(-0.198\pm0.074\right)\mu e^2] \: a_b,
\end{align}
where $e$ is the eccentricity of the binary orbit; $\mu$ is the mass ratio $M_2 / (M_1 + M_2)$; and $a_b$ is the semimajor axis of the pulsar's orbit. For the majority of the binary systems in this data set, $e \approx 0$, $M_{1} \approx$ 1.5 $M_{\odot}$, and $M_{2} \approx$ 0.2 $M_{\odot}$ ($\mu \approx 1/9$). Given these parameters, Equation \ref{eq:stype} reduces to
\begin{align}
\label{eq:stype_red}
a \lessapprox 0.4 a_b.
\end{align}
PSR J1903+0327 is the one notable exception to this approximation due to its eccentric orbit ($e \approx 0.44$) with a main-sequence star \citep{f+16}.

However, because of the relatively compact orbits of the binaries in the NANOGrav 11-year data set, the periods (listed in Table \ref{tab:psr_info}) corresponding to critical semimajor axes of planets in this configuration around a NANOGrav pulsar would be extremely short and therefore below the one-week limit set by NANOGrav's sampling frequency. We determine that a planet in an S-type configuration would not be detectable in the NANOGrav data set.

We then consider bodies in planetary P-type orbits, where the planet orbits both the pulsar and its companion as if they are a single mass. \citet{hw99} here define a lower limit for the semimajor axes of planets in this configuration, and thus possible values follow the relationship  
\begin{align}
\label{eq:ptype}
a \geq [&(1.60\pm0.04)+(5.10\pm0.05)e +\nonumber \\
&(-2.22\pm0.11)e^2 + (4.12\pm0.09)\mu +\nonumber \\
&(-4.27\pm0.17)e\mu + (-5.09\pm0.11)\mu^2 +\nonumber \\
&(4.61\pm0.36)e^2\mu^2] \: a_{b}.
\end{align}
As with Equation \ref{eq:stype}, using our assumptions that $e \approx 0$ and $\mu \approx 1/9$, we can reduce Equation \ref{eq:ptype} to
\begin{align}
\label{eq:ptype_red}
a \gtrapprox 2 a_{b}.
\end{align}
Due to the close orbits of the NANOGrav binaries, the majority of the minimum planetary periods in P-type configurations (also listed in Table \ref{tab:psr_info}) are smaller than the 7-day limit defined by our observation frequency. For these systems, we adopt 7 days as the lowest orbital period to probe; for systems with minimum periods that are greater than 7 days, we use the constraints set by their binary orbits to define the minimum orbital period to investigate.

We also investigate period constraints set by possible interactions with the timing models at lower frequencies. Certain patterns in pulsar timing data recur periodically as a result of Earth's rotation around the sun. In order to remove these systematics to better model the behavior of the pulsar, the timing models fit for the pulsar's position and proper motion, which ultimately erases power from any periodic signals with a frequency of 1 yr$^{-1}$, and for parallax, which will remove power at frequencies of 2 yr$^{-1}$. Though fitting for parallax does not result in a significant chance that signals with frequencies of 2 yr$^{-1}$ will go undetected, we find that it would be nearly impossible to detect signals of about 1 yr$^{-1}$ due to {\tt TEMPO}'s fitting procedure for position and proper motion. We also define a lower limit for potential orbital frequencies as $1/T$, where $T$ is the total span of the data. 

The NANOGrav data set timing models use a combination of white and red noise in the model of the timing of each pulsar. 

In the NANOGrav data set timing models, the white noise is modeled by three factors: EFAC, a multiplicative factor applied to measured TOA uncertainties; EQUAD, a factor added in quadrature to measured TOA uncertainties; and ECORR, a factor that accounts for noise correlated across frequencies in data collected across a wide band.  Such white noise models could hide the presence of signals from pulsar planets.  Therefore, for this paper, we excluded the white noise model (setting EFAC to 1 and EQUAD and ECORR to 0).  Theoretically, this reduction in TOA uncertainties could lower our detection threshold and produce spurious detections of planet signals.  In practice, as described below, we had no detections of planets, so we believe the reduction in TOA uncertainties does not contaminate our results.

NANOGrav also uses parameters to account for red noise. However, we only expect red noise to be significant on timescales as long as or longer than the data set, and we only search for planets with periods up to 2000 days, so the strong red noise exhibited by some of the pulsars in this data set is unlikely to affect our analysis. We therefore adopt the red noise parameters found by \cite{abb+18} and use ``unwhitened'' residuals, or residuals off of which the red-noise model has not been subtracted. We then rerun {\tt TEMPO} on all the pulsars in the 11-year data set to generate new timing models given these noise parameters. 

\section{Detection Techniques} \label{sec:techniques}

\begin{figure}
\begin{center}
\includegraphics[scale=.46]{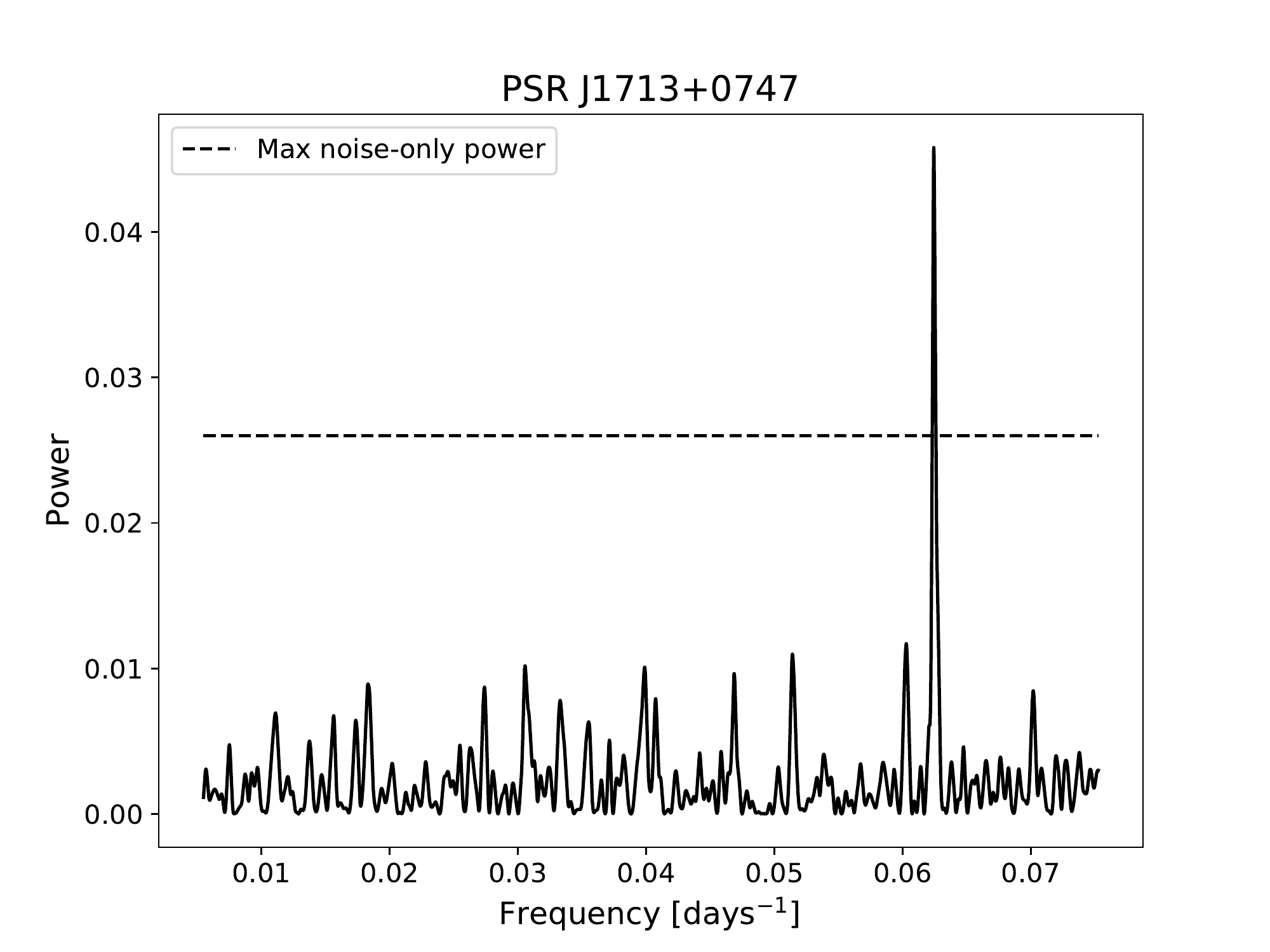}
\caption{\label{fig:J1713+0747}The resulting Lomb-Scargle periodogram of simulated residuals injected with a planetary signal for PSR J1713+0747. This power spectrum demonstrates a clear peak at a frequency of 0.062 days$^{-1}$ after being injected with a low-amplitude sinusoid at the lower limits of our detection capabilities. The dashed line represents the maximum power achieved through simulations of pure-noise residuals.}
\end{center}
\end{figure}

As discussed above, we would expect to see sinusoidal variation in a pulsar's timing residuals if a planet is present in the system. To detect these sinusoids, we use Astropy's {\tt LombScargle}\footnote{\url{https://docs.astropy.org/en/stable/api/astropy.timeseries.LombScargle.html}} module (named after \citet{l76} and \citet{s82}) to perform a periodogram analysis for detecting sinusoids in our data that may not be visible by eye \citep{apy13,vci+12,vi15}. The Lomb-Scargle periodogram is a tool specifically designed to be used with unevenly-sampled data with unequal errors bars to fit sinusoids of various frequencies to a data set. The quality of a sinusoid's fit to the data is determined by how dramatically the reduced-$\chi^2$ of the data improves if a fitted sinusoidal signal were removed. The Astropy module produces dimensionless, normalized power spectra that reflect these determinations with high spikes in power at the frequencies associated with the possible sinusoids identified in the data.

To use this module to detect potential planetary signals, we define a detection threshold based on the maximum power that could be achieved by residuals without a planetary signal. We simulate 10,000 realizations of white-noise residuals for each pulsar with each data point being taken from a Gaussian distribution centered at 0 that has a standard deviation equal to the error on that TOA. We also set the error on this simulated TOA equal to the error on the observed TOA. We set our detection threshold as the highest power value generated from the periodograms run on these simulated residuals. We then run the real residual data for each pulsar through the Astropy {\tt LombScargle} module and define a potential planetary detection as the presence of a power spike greater than the threshold value defined above.

We also use the {\tt LombScargle} module to set lower limits on the planetary masses that can be detected using pulsar timing by injecting planetary signals into simulated residuals. Because we cannot be sure no undetected planetary signal exists in the NANOGrav 11-year data, we again simulate residuals that mirror the noise levels of each pulsar but are guaranteed to contain only white noise. We define a frequency grid of 1,000 linearly-spaced frequencies with the maximum and minimum defined as in Section \ref{sec:sig} above. We then inject planetary signals with these frequencies and known amplitudes into the data and attempt to detect them using Lomb-Scargle periodograms and the detection technique described above. 

\newcolumntype{C}[1]{>{\centering\let\newline\\\arraybackslash\hspace{0pt}}m{#1}}

\begin{table}
\footnotesize
    \begin{tabular}{C{0.088\textwidth}  C{0.114\textwidth} C{0.1145\textwidth}    C{0.15\textwidth}}
        \hline
        Pulsar & P$_{\rm pl, max}$, S-type & P$_{\rm pl, min}$, P-type & M$_{\rm pl}$, P = 100 days  \\
            & (days) & (days) & ($M_{\rm moon}$) \\ 
        \hline
        
J0023+0923	&	2.58$\times$10$^{-5}$&	4.70$\times$10$^{-4}$	&	1.00	\\
J0030+0451	&	$-$	&	$-$	&	0.47	\\
J0340+4130	&	$-$	&	$-$	&	1.90	\\
J0613$-$0200	&	4.53$\times$10$^{-3}$&	0.08	&	0.28	\\
J0636+5128	&	$-$	&	$-$	&	1.14	\\
J0645+5158	&	$-$	&	$-$	&	0.91	\\
J0740+6620	&	$-$	&	$-$	&	1.45	\\
J0931$-$1902&	$-$	&	$-$	&	2.68	\\
J1012+5307	&	1.76$\times$10$^{-3}$&	0.03	&	0.87	\\
J1024$-$0719	&	$-$	&	$-$	&	1.08	\\
J1125+7819	&	$-$	&	$-$	&	3.64	\\
J1453+1902	&	$-$	&	$-$	&	8.39	\\
J1455$-$3330	&	0.73	&	13.30	&	2.19	\\
J1600$-$3053	&	0.10	&	1.89	&	0.22	\\
J1614$-$2230	&	0.15	&	2.74	&	0.38	\\
J1640+2224	&	1.63	&	29.77	&	0.67	\\
J1643$-$1224	&	0.50	&	9.08	&	0.26	\\
J1713+0747	&	0.73	&	13.29	&	0.16	\\
J1738+0333	&	7.99$\times$10$^{-4}$	&	0.01	&	1.09	\\
J1741+1351	&	0.14	&	2.64	&	0.44	\\
J1744$-$1134	&	$-$	&	$-$	&	0.56	\\
J1747$-$4036	&	$-$	&	$-$	&	1.44	\\
J1832$-$0836	&	$-$	&	$-$	&	1.00	\\
J1853+1303	&	1.03	&	18.80	&	0.97	\\
B1855+09&	0.11	&	2.03	&	0.21	\\
J1903+0327	&	1.51	&	137.84	&	0.32	\\
J1909$-$3744	&	0.01	&	0.19	&	0.13	\\
J1910+1256	&	0.39	&	7.02	&	0.53	\\
J1911+1347	&	$-$	&	$-$	&	1.10	\\
J1918$-$0642	&	0.10	&	1.74	&	0.72	\\
J1923+2515	&	$-$	&	$-$	&	1.33	\\
B1937+21	&	$-$	&	$-$	&	0.01	\\
J1944+0907	&	$-$	&	$-$	&	0.93	\\
B1953+29	&	0.70	&	12.72	&	0.88	\\
J2010$-$1323	&	$-$	&	$-$	&	0.85	\\
J2017+0603	&	0.01	&	0.23	&	0.54	\\
J2033+1734	&	$-$	&	$-$	&	1.45	\\
J2043+1711	&	0.01	&	0.15	&	0.27	\\
J2145$-$0750	&	0.13	&	2.34	&	1.53	\\
J2214+3000	&	5.70$\times$10$^{-5}$	&	1.04$\times$10$^{-3}$	&	1.98	\\
J2229+2643	&	$-$	&	$-$	&	1.84	\\
J2234+0611	&	$-$	&	$-$	&	1.30	\\
J2234+0944	&	$-$	&	$-$	&	1.51	\\
J2302+4442	&	1.46	&	26.67	&	1.90	\\
J2317+1439	&	0.01	&	0.25	&	0.45	\\

    \end{tabular}
    
    \caption{Orbital characteristics of planets around binary MSPs and mass limits for planets with P=100 days. P$_{\rm pl,max}$ is the maximum orbital period possible for planets in an S-type configuration. P$_{\rm pl,min}$ is the minimum orbital period of planets in P-type configurations. M$_{\rm pl}$ is the minimum planetary mass detectable in a circular orbit using the 11-year data set given a 90\% confidence level and adjusted by the factor of 2.5 discussed in Section \ref{sec:results}. }
    \label{tab:psr_info}
\end{table}

At each chosen frequency, we lower the amplitude of the injected sinusoid until it can no longer be detected more than 90\% of the time over 1,000 simulations of noisy residuals. Figure \ref{fig:J1713+0747} shows an example of simulated residuals injected with a planetary signal with an amplitude at the lower limit of our detectable range. If a periodogram spike is detected with a value greater than our noise-only threshold, we also identify whether the periodogram has recovered the correct period of the injected signal by determining whether the number of planetary cycles  of the recovered signal over the span of the data set matches that of the injected signal. After using the amplitude and frequency data to determine the lowest detectable mass for each orbital period, we perform a linear, least-squares fit in logarithmic space to obtain a best-fit line for each pulsar's mass-period relationship (see Figure \ref{fig:planet lims}). 

To ensure that our method for determining lower limits is realistic, we also run {\tt TEMPO} on pulsar residuals that include an injected planetary signal, to see if covariances with other timing parameters will affect planet detectability. Due to the extremely time-consuming nature of running thousands of {\tt TEMPO} iterations, we choose to test the validity of our detection scheme with a single simulated pulsar under the assumption that the results will be reflective of all the pulsars in this data set. We simulate pulsar TOAs spanning about 4600 days that reflect NANOGrav's uneven sampling rate and have uniform 1 $\mu$s errors, making it a fairly typical NANOGrav MSP, and we add a sinusoid to the TOAs that is representative of a planetary signal. We then run {\tt TEMPO} again to obtain a model for the simulated pulsar with its injected signal and proceed to attempt to detect the signal as described above. 

\section{Results} \label{sec:results}

\begin{figure}
\begin{center}
\includegraphics[scale=0.42]{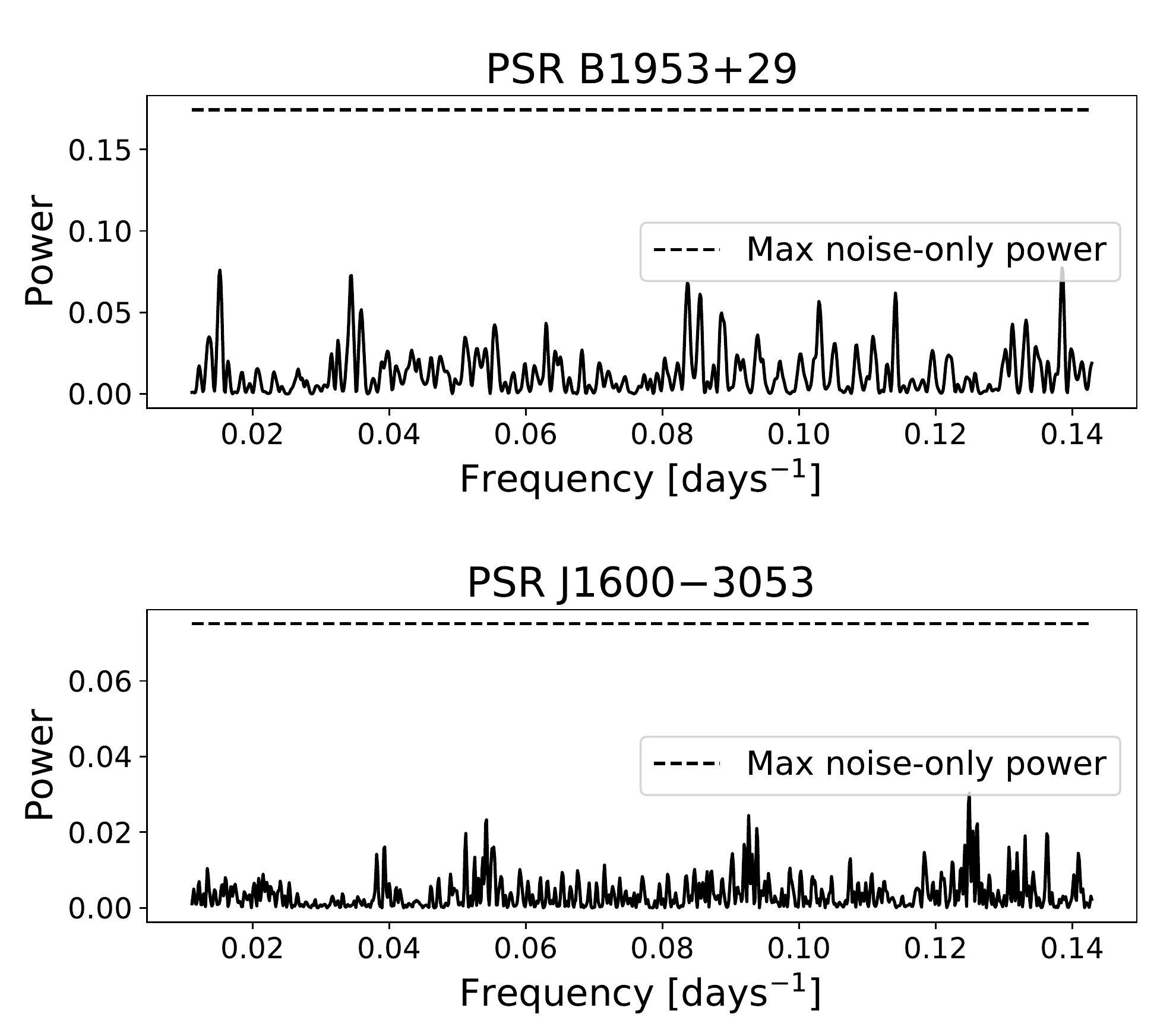}
\caption{\label{fig:non_detect}The results of Lomb-Scargle applications to PSR B1953+29 and PSR J1600$-$3053. The power spectra for these pulsars show no significant peaks in power over the frequencies investigated. The dashed lines again represent the maximum power level reached in simulations of noise-only residuals, indicating that no planetary signals exist in these residuals.}
\end{center}
\end{figure}

No exoplanets are discovered around the pulsars in this data set. Figure \ref{fig:non_detect} shows two examples of power spectra from the Lomb-Scargle periodograms. As shown, no significant peaks stand out in the spectra, and all power values are well below the thresholds determined for each pulsar. We therefore conclude that no planets orbiting with periods within our range of sensitivity exist around the pulsars in the NANOGrav data set.

\begin{figure*}
\begin{center}
\includegraphics[scale=0.7]{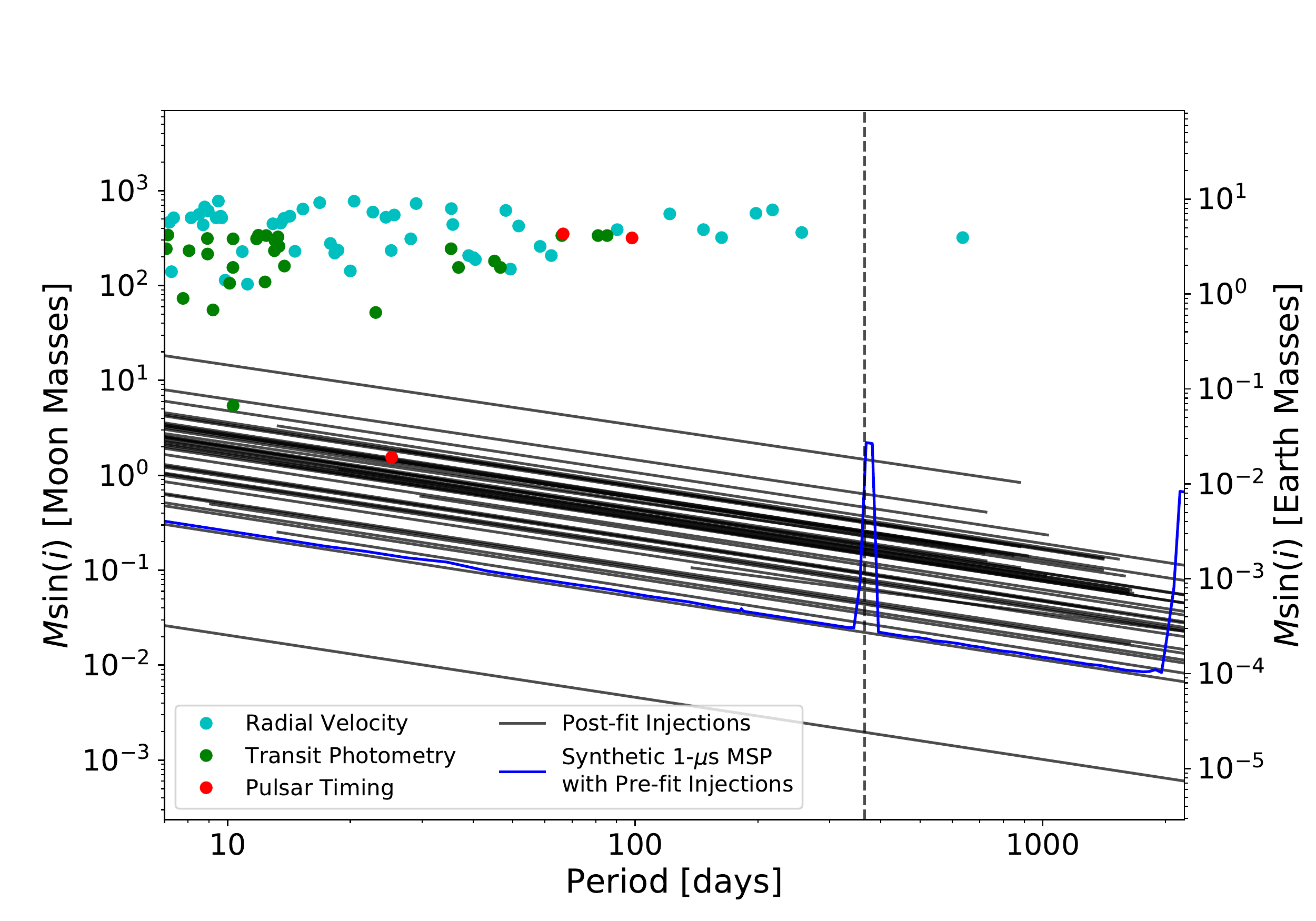}
\caption{\label{fig:planet lims}The lower limits of detectable masses in the 11-year NANOGrav data set. The solid black lines show the linear fits for the lowest detectable planetary masses using our original method where planetary signals are injected post-{\tt TEMPO} fitting. The slopes of those lines are due to Kepler's laws.  As a result of our simulations of a pulsar with planetary signals injected pre-{\tt TEMPO} fitting, the lines were adjusted upward by a factor of 2.5 to reflect the overall sensitivity loss that would occur from running {\tt TEMPO}'s fitting procedure on pulsar data already containing a planetary signal. The blue line shows the mass-period relationship derived from running {\tt TEMPO} on simulated pulsar data containing a planetary signal and with average white noise of 1-$\mu$s. The results of the simulated pulsar support the validity of our original method injecting planetary signals post-fitting but indicate that {\tt TEMPO} fitting procedures cause an overall sensitivity loss that increases the mass limits by a factor of 2.5 and results in additional sensitivity loss at a period of 1 year (indicated by the dashed line) and at periods greater than $\sim$2000 days. The colored data points represent the least massive 10\% of exoplanets discovered using alternate methods.}
\end{center}
\end{figure*}

The results of the simulated planetary signal injections reinforce our conclusion regarding the presence of exoplanets around these pulsars. The injections demonstrate the incredible sensitivity of NANOGrav pulsar timing data to planetary perturbations, which is shown in Figure \ref{fig:planet lims}, with the mass constraints for planets with periods of 100 days listed in Table \ref{tab:psr_info}. Within the range of orbital periods investigated here, all of the NANOGrav pulsars demonstrate a sensitivity to planetary masses well below an earth mass and even as low as a fraction of a moon mass. These limits also demonstrate a -2/3 slope as a function of orbital period, which results from Equation \ref{eq:Kep3} as well as NANOGrav's roughly equal sensitivity to planetary perturbations across frequencies.

The detections made after running {\tt TEMPO} on simulated residuals with an injected planetary signal are also shown in Figure \ref{fig:planet lims}. There is a significant loss of sensitivity at 1 year due to fitting for pulsar position and proper motion. Our ability to detect planetary signals wanes further at periods of about 2000 days or greater when the injected signal's period becomes equal to or longer than the range of the data set. 

We also find that running {\tt TEMPO} on data that already contains a planetary signal, yet with no planets in the timing model, will result in a small loss of sensitivity. {\tt TEMPO's} model fitting procedure with the incorrect timing model absorbs some of that planetary signal, causing us to become less sensitive to planets. We used synthetic TOAs from an isolated pulsar with a constant DM to calibrate the mass-detection penalty and find it to be approximately a factor of 2.5 over the planetary orbital periods of interest (i.e., when searching {\tt TEMPO}-derived residuals for an injected signal in the pulse arrival times, the planets we are able to detect are ~2.5 times more massive than those whose signals are injected directly into the residuals themselves). As a result, we increase the lower limits found using our original method by this factor of 2.5 in order to more accurately reflect the lowest planetary masses that would be detected after using {\tt TEMPO} to fit and model the NANOGrav pulsars. Additionally, since no red noise parameters are used to model the noise of the simulated pulsar, the post-{\tt TEMPO} simulated data limit appears slightly more sensitive than the constraints for the real pulsars, whose models account for red noise, after they are adjusted by a factor of 2.5. 

Figure \ref{fig:planet lims} also indicates how the sensitivity of the NANOGrav pulsars compares with that of other exoplanet detection methods. Using data from the NASA Exoplanet Archive, the least massive 10\% of exoplanets found using the listed methods are plotted over the mass limits we find using timing of the NANOGrav pulsars. In general, the NANOGrav pulsars demonstrate that pulsar timing is far more sensitive to planetary detections than other methods, sometimes by orders of magnitude.

\vspace{10mm}

\section{Conclusion} \label{sec:conc}
As a result of the exquisite sensitivity stemming from the high timing precision and long baselines of the NANOGrav data set, we determine that pulsar timing is capable of detecting planetary masses orders of magnitude smaller than any other detection method. Our results are primarily limited by the sampling frequency of the NANOGrav data.  In Section \ref{sec:sig}, our investigations show that there is a physical possibility of planets existing with orbital periods shorter than those that can be detected given this data set's sampling frequency. More closely spaced timing observations, taken several times a day over the span of weeks or months, would allow us to probe higher-frequency orbits to which we are not currently sensitive. It is worth noting, however, that if such short-period planets existed in our data with fairly massive planetary companions, the white-noise parameters from our standard NANOGrav fits would indicate substantial excess noise in our data (i.e. via large EFACs, for instance), which is something that we do not presently see.

We are of course also limited by the frequency with which planets exist around pulsars, which is not currently known. Since only a handful of planetary-mass bodies have to date been discovered around three or four pulsars, it seems likely that their existence around millisecond pulsars is quite rare. Further investigations are needed to understand the demographics of this population of planets. 

\bigskip

The NANOGrav project receives support from National Science Foundation (NSF) Physics Frontiers Center award number 1430284. The National Radio Astronomy Observatory and the Green Bank Observatory are facilities of the National Science Foundation operated under cooperative agreement by Associated Universities, Inc. The Arecibo Observatory is a facility of the National Science Foundation operated under cooperative agreement (\#AST-1744119) by the University of Central Florida in alliance with Universidad Ana G. Méndez (UAGM) and Yang Enterprises (YEI), Inc. Portions of this work performed at NRL are supported by the Chief of Naval Research. SMR is a CIFAR Fellow. The NASA Exoplanet Achive is operated by the California Institute of Technology under contract with the National Aeronautics and Space Administration.

\bibliographystyle{aasjournal}
\bibliography{pulsarPlanets}
\end{document}